\documentclass[aps,prd,amssymb,superscriptaddress,nofootinbib,twocolumn,preprintnumbers]{revtex4-1} 	
\usepackage[utf8]{inputenc}
\usepackage[english]{babel}
\usepackage{amsmath}
\usepackage{amsfonts}
\usepackage{amssymb}
\usepackage{amsthm}
\usepackage{color}
\usepackage{graphicx}
\usepackage[force]{feynmp-auto}
\usepackage{feynmp}
\usepackage[hidelinks]{hyperref}

\newcommand\bw{\begin{widetext}}
\newcommand\ew{\end{widetext}}

 \def\be{\begin{equation}}
\def\ee{\end{equation}}
 \def\ba{\begin{align}}
\def\ea{\end{align}}
\def\bea{\begin{eqnarray}}
\def\eea{\end{eqnarray}}

\def\di{\mathrm{d}}

\def\b{\beta}
\def\g{\gamma}

\def\l{\lambda}

\def\r{\rho}

\def\r{ř}

\begin{document}
\preprint{CERN-TH-2017-100, INR-TH-2017-011, FR-PHENO-2017-015}
\title{Ho\r ava gravity is asymptotically free (in 2+1 dimensions)}

\author{Andrei O.~Barvinsky}\affiliation{Theory Department, Lebedev
  Physics Institute, Leninsky Prospect 53, Moscow 119991, Russia}
  \affiliation{Tomsk State University, Department of Physics, Lenin Ave. 36, Tomsk 634050, 
Russia}

\author{Diego Blas}\affiliation{Theoretical Physics Department, CERN,
CH-1211 Geneva 23, Switzerland}

\author{Mario Herrero-Valea}\affiliation{Institute of Physics, LPPC,
  Ecole Polytechnique F\'ed\'erale de Lausanne, CH-1015, Lausanne,
  Switzerland} 

\author{Sergey M.~Sibiryakov}
\affiliation{Theoretical Physics Department, CERN,
CH-1211 Geneva 23, Switzerland}
\affiliation{Institute of Physics, LPPC,
  Ecole Polytechnique F\'ed\'erale de Lausanne, CH-1015, Lausanne, Switzerland} 
\affiliation{Institute for Nuclear Research of the
Russian Academy of Sciences, 60th October Anniversary Prospect, 7a, 117312
Moscow, Russia}

\author{Christian F.~Steinwachs}
\affiliation{Physikalisches Institut, Albert-Ludwigs-Universit\"at 
Freiburg, Hermann-Herder-Strasse 3, 79104 Freiburg, Germany}

\begin{abstract}
We compute the $\b$-functions of marginal couplings in projectable
Ho\v rava gravity in $2+1$ spacetime dimensions. 
We show that the renormalization group flow
has an asymptotically-free fixed point in the ultraviolet (UV),
establishing the theory as a UV-complete model with dynamical
gravitational degrees 
of freedom. Therefore, this theory may serve as a toy-model to study
fundamental aspects of quantum gravity. Our results represent a step forward
towards understanding the UV properties of realistic versions of Ho\v rava
gravity. 
\end{abstract}

\maketitle

\paragraph*{{\bf Introduction} -}
Formulating a quantum theory of the gravitational interactions
remains one of the major challenges in theoretical physics. Despite
the impressive achievements in this direction, in particular of string
theory, many fundamental questions still remain open. It is of
particular interest whether such a theory can be formulated in the
language of a unitary and perturbative quantum field theory in $3+1$
space-time dimensions, as it is the case for all the other
interactions in the Standard Model of particle physics (SM).  

The main obstruction for this venture within general relativity (GR) 
is the well-known fact that the gravitational coupling constant, the
Newton's constant $G$, is dimensionful for dimensions greater than
$1+1$. This ultimately renders the theory perturbatively
non-renormalizable -- with an increasing number of divergences
appearing at every order in the loop expansion. A possible solution is
to extend the gravitational Lagrangian by terms quadratic in the
curvature, thereby increasing the number of derivatives acting on the
metric field. 
Although this can make the
theory renormalizable \cite{Stelle:1976gc} and even asymptotically
free \cite{Fradkin:1981hx,Avramidi:1985ki}, it jeopardizes the
unitarity of the theory. 

This idea was revisited by
P. Ho\r ava, who suggested that unitarity and perturbative
renormalizability could be reconciled in theories of gravity without
Lorentz invariance (LI) \cite{Horava:2009uw}. More concretely, he
suggested that if the universe is endowed with a preferred foliation
into space and time, one can construct a power-counting renormalizable
theory containing only  
marginal and relevant operators with 
respect to a Lifshitz (anisotropic) scaling 
\begin{align}\label{eq:lifshitz_scaling}
t\mapsto b^{-d} t,\qquad x^i \mapsto  b^{-1}\, x^i,
\end{align}
where $d$ is the number of spatial dimensions. 
This proposal requires a mechanism to account for the stringent tests
of LI of the SM and some promising ideas have been suggested, see
\cite{Liberati:2013xla}. It is important to stress that so far
`non-projectable’ Ho\v rava gravity (defined below) is consistent with
all phenomenological constraints \cite{Blas:2009qj,Blas:2014aca}. 
Finally, apart from being a viable candidate for the ultraviolet (UV)
completion of GR, Ho\v rava gravity can be used as a gravitational
dual for strongly coupled systems exhibiting the Lifshitz scaling
\cite{Janiszewski:2012nb,Griffin:2012qx}.

Several fundamental aspects of Ho\v rava gravity must be clarified 
before declaring it a successful theory of quantum gravity. The first
one is renormalizability beyond power counting. Recently, this has
been 
proven for the projectable version (defined below)
\cite{Barvinsky:2015kil,Barvinsky:2017zlx}, whereas the non-projectable case 
still remains elusive. 
Second, the ultraviolet structure of the theory, essential to
establish its
consistency, is not known beyond tree-level. Partial results
include the study of renormalization group (RG) flow
in a simplified model obtained from the projectable theory
 by a conformal reduction of the metric \cite{Benedetti:2013pya} and
 the calculation of the contributions of matter loops to the
RG running of the couplings \cite{DOdorico:2014tyh}.

In this {\it letter} we make the next step in this direction and
compute the complete $\beta$-functions in the pure
projectable Ho\v rava gravity in $2+1$ dimensions, 
for the first time fully taking into account the gravitational degrees of freedom. It is worth
noting that, unlike $(2+1)$-dimensional GR, the theory possesses  a
local propagating scalar mode \cite{Sotiriou:2011dr}. We find that there is a region in
parameter space where the projectable theory is asymptotically
free. This implies that projectable Ho\r ava
gravity is a perturbatively complete theory of quantum gravity in
$2+1$ dimensions.   

\paragraph*{{\bf Ho\v rava gravity (projectable case)} -}

The action of Ho\v rava gravity in $d+1$ spacetime 
dimensions is constructed by considering theories invariant under the 
diffeomorphisms compatible with a preferred foliation of space-time 
\be
\label{eq:fdiff}
t\mapsto {t}'(t),\qquad x^i \mapsto {x'}^i(x,t). 
\ee
This suggests to decompose the metric in the Arnowitt--Deser--Misner (ADM) 
form\footnote{We perform our calculation in Euclidean signature.}
\be
\di s^2=N^2 \di t^2+\g_{ij}(\di x^i +N^i \di t)(\di x^j +N^j \di t),
\ee
$i,j=1,2,...,d$. Here $N$ is the lapse function, $N^i$ the shift
vector and $\gamma_{ij}$ the metric on the spatial co\-di\-men\-sion-one
hypersurfaces which foliate the spacetime. 
The gravitational action with only marginal and relevant couplings
with respect to the Lifshitz scaling \eqref{eq:lifshitz_scaling} reads, 
\be
S=\frac{1}{2G}\int \di t \,\di ^dx\;N\sqrt{\g}\;\left(K_{ij}K^{ij}-\lambda K^2+{\cal V}\right),\label{eq:action_g}
\ee
where $\lambda$ and $G$ are coupling constants while $K_{ij}$ is the
extrinsic curvature of the foliation  
\be\label{eq:K}
K_{ij}=\frac{1}{2N}\left(\partial_t \g_{ij}-\nabla_{i}N_j-\nabla_j N_i\right),
\ee
with $\nabla_i$ being the covariant derivative compatible with
$\gamma_{ij}$ and $K\equiv \gamma^{ij}K_{ij}$. The potential
${\cal V}$ includes all possible terms constructed out of the covariant
derivative $\nabla_i$, the spatial Riemann tensor $R_{ijkl}$ and the
acceleration vector $a_i=\partial_i \log N$, that under
\eqref{eq:lifshitz_scaling}  
have scaling dimensions $2d$ or less. The action
\eqref{eq:action_g} is then power-counting renormalizable
\cite{Horava:2009uw}. The case of GR would correspond to
$\lambda=1$ and ${\cal V}=2\Lambda-R$, where $\Lambda$ is the
cosmological constant and $R$ is the intrinsic curvature of the
$d$-dimensional slices.

The symmetry \eqref{eq:fdiff} can be satisfied if $N$ depends only on
time ($a_i=0$), which defines the {\it projectable} version of Ho\v
rava gravity. 
The action for this proposal was derived in \cite{Sotiriou:2009gy,Sotiriou:2009bx}
and it was shown to be renormalizable beyond power counting in
\cite{Barvinsky:2015kil,Barvinsky:2017zlx}. This proposal is the focus of this work. We also
concentrate only on the case $d=2$. The previous restrictions reduce
the number of terms in $\cal V$ drastically, while keeping non-trivial
(gravitational) local degrees of freedom \cite{Sotiriou:2011dr}. In the
projectable case, the lapse function $N$ does not affect
the local dynamics. By fixing the time coordinate it can be set to
$N=1$ in perturbation theory, which we will assume in what
follows. 
The remaining gauge
invariance consists of time-dependent spatial diffeomorphisms. An
important caveat is that this theory is not phenomenologically viable
in $3+1$ dimensions \cite{Blas:2010hb}. Nevertheless, one can view it
as a toy model for quantum gravity with dynamical degrees
of freedom.

\paragraph*{{\bf Quantizing projectable Ho\v rava gravity in $2+1$} -}

For $d=2$ the potential term of projectable Ho\r ava gravity has the form,
\be\label{eq:action_2}
{\cal V}=2\Lambda+ \mu R^2.
\ee
The term linear in $R$ is not present, as it is a total derivative in $2+1$ dimensions. Our aim is to perform a one-loop calculation and to analyse the running of the marginal couplings that define the theory. We do this by using the background field technique \cite{Abbott:1981ke}, where the renormalization of the coupling constants is captured by the contributions to operators of the background fields, coming from integrating out the quantum fluctuations around the background. 
The anomalous dimension of the cosmological constant was already
computed in \cite{Griffin:2017wvh}. We focus on the case $\Lambda=0$
and study the renormalization group (RG) running of the three marginal
couplings $\{G,\lambda, \mu  \}$ governing the UV dynamics of the
theory. 
Only two combinations
$\left\{{\cal{G}}\equiv G/\sqrt{\mu},\lambda\right\}$ are physical
in the sense that their $\beta$-functions do not depend on the choice
of the gauge. Conversely, all gauge-invariant quantities depend only
on these combinations. This is a consequence of the fact that a change
of gauge shifts the off-shell effective action by a contribution that
vanishes on-shell \cite{DeWitt:1967ub,Kallosh:1974yh}. The equations of motion
following from \eqref{eq:action_g} and \eqref{eq:action_2} imply the
global Hamiltonian constraint, 
\be
{\cal H}\equiv \int \di ^2 x\; \sqrt{\g}\;\left[K_{ij}   K^{ij}-\lambda   K^2 - \mu  R^2 \right]=0.
\ee
This is also a unique combination of marginal gauge-invariant operators vanishing on-shell. 
Thus, the 1-loop effective action $\Gamma[ \gamma_{ij}, N_i]$ is
defined up to the transformations,
\be
\Gamma\rightarrow \Gamma+\varepsilon \int \di t \di ^2 x\; \sqrt{\g}\;\left[ K_{ij}K^{ij}-\lambda K^2 - \mu R^2 \right],
\ee
with arbitrary constant $\varepsilon$. 
Such a shift corresponds to the following transformation of
couplings 
$\delta G=-2\,G^2\,\varepsilon$, $\delta \lambda=0$ and $\delta
\mu=-4\,G\,\mu\,\varepsilon$. 
Only the $\beta$-functions of the {\it essential} couplings $\lambda$
and $\cal G$, which are invariant under these transformations, have
physical meaning. Here $\lambda$ measures the
deviation from a relativistic invariant kinetic term, while $\cal G$
controls the strength of 
gravitational interactions. 

In order to compute $\Gamma[\gamma_{ij}, N_i]$, we expand the fields
around an arbitrary background configuration 
\begin{align}
\gamma_{ij}=\bar{\gamma}_{ij}+h_{ij},\qquad N_i=0+n_i,
\end{align}
where we have set $\bar{N}_i=0$. This choice simplifies the
calculation without losing information about the $\beta$ functions,
since $\bar{N}_i$ only enters via $ \bar{K}_{ij}$. Expanding the bare
action (\ref{eq:action_g})
up to second order in the quantum fluctuations yields the quadratic
action $S_2$, which is
sufficient to capture the one-loop contributions. For notational
simplicity, we omit the bars over background quantities in what
follows.  

We add to $S_2$ a gauge fixing term implementing the regular gauge proposed in \cite{Barvinsky:2015kil}, 
\begin{align}\label{eq:gauge_fixing}
S_{\rm gf}=\frac{\sigma}{2G}\int \di t \di^2x\; \sqrt{\gamma}\;F_i\frac{-1}{\gamma_{ij}\Delta+\xi\nabla_i \nabla_j}F_{j},
\end{align}
with the gauge fixing condition $F_{i}$ given by
\begin{align}
&F_i=\partial_t n_i-\frac{1}{2\sigma}(\gamma_{ij}\Delta+\xi \nabla_i \nabla_j)(\nabla^k h_{k}^{j}-{\lambda} \nabla^j h).
\end{align}
This gauge fixing leads to regular propagators featuring a
uniform scaling structure, which is a key element in the proof of the
renormalizability of the theory
\cite{Anselmi:2008bq,Barvinsky:2015kil}. Moreover, this choice of
$F_i$ removes the mixing between $h_{ij}$ and $n_i$ in the quadratic
part of the action. All the terms in $S_{\rm gf}$ are local -- except for
one, which, however, can be written in a local form by ``integrating
in'' the extra auxiliary field $\pi^i$ \cite{Barvinsky:2015kil}, 
\begin{align}
&\frac{\sigma}{2G} \int \di t \di^2x\sqrt{\gamma}\;\partial_t n_i \frac{-1}{\gamma_{ij}\Delta+\xi \nabla_i \nabla_j} \partial_t n_j \mapsto\\
&\frac{1}{2G}\int \di t \di^2x\sqrt{\gamma}\;\left( \frac{-1}{2\sigma}\pi^i (\gamma_{ij}\Delta+\xi \nabla_i \nabla_j) \pi^j -i \pi^i \partial_t n_i\right). \nonumber
\end{align}
Finally, we need to include the corresponding ghost action, which is
constructed in the standard way from the gauge fixing condition $F_i$  
\cite{Barvinsky:2015kil,Griffin:2017wvh}, 
\begin{align}\label{eq:ghost_action}
\nonumber S_{\rm gh}
=&-\int \di t\di^2x\sqrt{\gamma}\bar{c}^i \bigg[ \partial_t \left(\gamma_{ij}\partial_t c^j\right)-\frac{1}{2\sigma}\Delta^2 (\gamma_{ij}c^j)\\
\nonumber &-\frac{1}{2\sigma}\Delta \nabla_k \nabla_i c^k+\frac{\lambda}{\sigma}\Delta \nabla_i \nabla_j c^j-\frac{\xi}{2\sigma}\left( \nabla_i \nabla_j \Delta c^j\right. \\
&\left.+ \nabla_i \nabla_j\nabla_k\nabla^j c^k-2\lambda \nabla_i \Delta \nabla_j c^j   \right)\bigg].
\end{align}

Given the total action for the quantum fluctuations $S_{\rm
  tot}=S_{\rm 2}+S_{\rm gf}+S_{\rm gh}$, we use perturbation theory
around a background metric $\gamma_{ij}$ which is close to flat space
$\delta_{ij}$, 
\begin{align}
\label{backexpand}
\gamma_{ij}=\delta_{ij}+H_{ij}. 
\end{align}
Due to the invariance of the effective action with respect to the
background diffeomorphisms, it is sufficient to focus on the
renormalization of the terms
quadratic in $H_{ij}$. The bare action quadratic in
$H_{ij}$ reads 
\begin{align} 
\label{eq:action_H}
&S_{\rm H}=\frac{1}{2G}\int \di t\di^2x\bigg\{\tfrac{1}{4} \left(\dot{H}_{ij} \dot{H}^{ij} -   \lambda \dot{H}  \dot{H}\right)  \\
&- \mu \partial_{b}\partial_{a}H^{ab} (2\Delta H - \partial_{j}\partial_{i}H^{ij})   +\mu \Delta H  \Delta H +O(H^3)  \bigg\}.\nonumber
\end{align}
The $\beta$-functions are found by studying how the two-point
functions of $H_{ij}$, following from \eqref{eq:action_H}, 
are renormalized after integrating out the quantum fluctuations. 
The renormalization of $G$ is then extracted from
$\dot{H}_{ij}\dot{H}^{ij}$, while the one of $\lambda$ comes from
$\dot{H} \dot{H}$. For the renormalization of $\mu$ we can use any of
the three other  structures in $ S_{\rm H}$. 

\begin{figure}
\begin{center}
\includegraphics[scale=1]{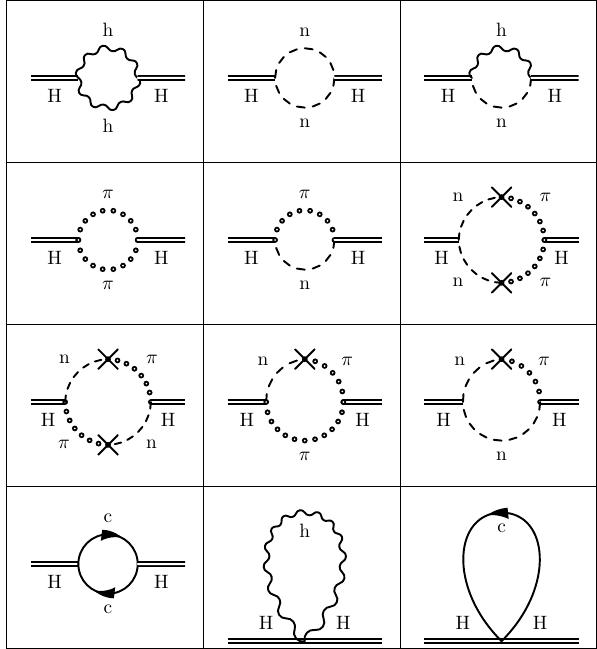} 
\end{center}
\caption{Feynman diagrams (bubbles and fishes) for the two point function of $H_{ij}$. The cross represents the mixed propagator $\langle n^i\pi^j\rangle $.}\label{fig:diagrams}
\end{figure}

The background independent part of the second order action defines the
propagators for $h_{ij}$, $n_i$, $\pi_i$, $\bar{c}^i$ and $c^i$.  
The latter are particularly simple in the gauge
\begin{align}\label{eq:sigmaxi}
\sigma=\frac{1 - 2 \lambda}{8 \mu (1 - \lambda)}, \qquad \xi=-\frac{1 - 2 \lambda}{2 (1 - \lambda)},
\end{align}
where they are all proportional to the propagator of the physical
scalar mode,
\begin{equation}
 \begin{split}
&{\cal P}_{\rm s}(\omega,p)=\left[\omega^2 +4\mu \frac{1-\lambda}{1-2\lambda}\;p^4\right]^{-1}. \label{eq:poles}
\end{split}
\end{equation}
The vertices required for the one-loop calculation can be found by
expanding $S_{\rm tot}$ up to second order in the background field
$H_{ij}$. The diagrams 
which give rise to logarithmic divergences are
shown in Fig.~\ref{fig:diagrams}.

We stress that the possibility to study only the 
renormalization of terms quadratic in $H_{ij}$ essentially relies on
our procedure 
where we start from the action for
perturbations $S_{\rm tot}$ 
invariant under the background gauge transformations, 
including the gauge fixing and
ghost terms. Only after that we expand the background according to
(\ref{backexpand}). Had we started by first expanding the metric
around flat spacetime and then fixing a (non-covariant) gauge, we
would have to compute also the renormalization of 3- and 4-point
vertices to factor out the (gauge-dependent) wavefunction
renormalization.

\paragraph*{{\bf Calculation of diagrams} -}
Although the number of diagrams in Fig.~\ref{fig:diagrams} is not so
large, the different vertices contain multiple terms which make
computations rather involved and lengthy. To handle this complexity we
use the Mathematica package \emph{xAct} \cite{Brizuela:2008ra} to
manipulate the algebraic expressions and FORM \cite{Ueda:2014sya} to
reduce the output of the diagrams. 
The computation is simplified by considering the renormalization of 
$\{{ G},\lambda\}$ and of $\mu$ separately. 
This can be extracted by evaluating the quadratic part of the
effective action $\Gamma[H_{ij}]$ on 
time- or space-dependent backgrounds which correspond 
respectively to diagrams with
vanishing spatial momenta or frequency in external
legs. 
Thus, for $\{{ G},\lambda\}$-renormalization we first focus on computing the 
contributions carrying only external frequency $\Omega$ at vanishing external
momentum $P_i$. Then we do the opposite for $\mu$-renormalization and 
compute the logarithmically divergent diagrams carrying only external 
momentum at vanishing $\Omega$.  

The prototypical loop integral over internal momentum and frequency 
has the form,
\begin{align}
\label{eq:inte}
\int \frac{\di\omega  \di^2q}{(2\pi)^3} \; \omega^{2a} q^{2b}\prod_I {\cal P}_{s} (\omega+\Omega_I,\,q+P_I),
\end{align}
with constant exponents $a$, $b$. Here $\Omega_I$ and $P_I$ are the
relevant external frequencies and two-momenta.
Since we are interested in the logarithmic divergences, we retain only
the contributions proportional to $\Omega^2$ or $P^4$, which renormalize the
terms shown in (\ref{eq:action_H}).
For this, we Taylor expand
the integrand of \eqref{eq:inte} up to the desired order in external
frequency or momentum, such that the  
final integrands all acquire the general form
\begin{align}
\label{eq:integrand}
{\cal I}[a,b,A]=\omega^{2a} q^{2b} ({\cal P}_s(\omega,q))^A ,
\end{align}
with $A$ being a constant power. 
We regularize the UV divergences, which appear in the integration of
\eqref{eq:integrand}, by using the
Schwinger integral representation for the propagator. 
To this end, we introduce an auxiliary
``proper time'' parameter $s$ and rewrite \eqref{eq:integrand} as 
\be
 {\cal I}[a,b,A]=\omega^{2a} q^{2b}\int_0^\infty \frac{\di s \; s^{A-1} }{\Gamma(A)}  \; e^{-s ({\cal P}_s(\omega,q))^{-1} }.
\ee
The integral over frequency and momentum in \eqref{eq:inte} can then
be expressed in terms of the
$\Gamma$-function.  
In the Schwinger integral representation UV divergences appear in the
limit $s\rightarrow 0$, with logarithmic divergences corresponding to
$\int \di(\log s)$.

\paragraph*{{\bf One-loop $\beta$-functions and asymptotic freedom} -}
In the Wilsonian picture the quantum corrections are interpreted as
the result of integrating out virtual modes with momenta between a
certain UV cutoff $\Lambda_{\rm UV}$ and the subtraction point $k_*$. 
This leads to the following identification in the divergent part of
the effective action,
\begin{align} 
\int\di (\log s)\mapsto\log\left(\frac{\Lambda_{\rm UV}^4}{k_*^4}\right)\;,
 \end{align}
where we have taken into account that the
parameter $s$ has scaling dimension $4$.  
The RG flow originates from the sensitivity of the couplings to a
change in $k_*$, see e.g. \cite{Liao:1994fp}. 
We find the following $\beta$-functions for the physical couplings:
\begin{subequations}
\label{eq:betas}
\begin{align}
&\beta_{\lambda}\equiv
 k_*
\frac{\di {\lambda}}{\di k_*}
=\frac{15-14\lambda}{64\pi}\sqrt{\frac{1-2\lambda}{1-\lambda}}\; {\cal G}\label{eq:betaoflambda},\\
&\beta_{\cal G}\equiv
 k_*
\frac{\di {\cal G}}{\di k_*}
=-\frac{(16-33\lambda +18\lambda^2)}{64\pi (1-\lambda)^2}\sqrt{\frac{1-\lambda}{1-2\lambda}}\; {\cal G}^2 .\label{eq:betaofg}
\end{align}
\end{subequations}
We have checked this result by performing
independent calculations in several gauges other than
(\ref{eq:sigmaxi}). Namely, we have considered the gauge
(\ref{eq:gauge_fixing}) with $\xi=0$ (and with $\sigma$ as in (\ref{eq:sigmaxi}))
and, 
outside of the family (\ref{eq:gauge_fixing}), the conformal gauge
$h_{ij}=e^{2\phi}\g_{ij}$ which is possible in two spatial
dimensions \cite{future}. Furthermore, the (gauge-dependent)
$\beta$-function for the coupling $\mu$ can be extracted from the results of
\cite{Griffin:2017wvh} and it agrees with our results when evaluated
in the same gauge. 

\begin{figure}[h]
\vspace{0.2cm}
\centerline{\includegraphics[width=0.36\textwidth]{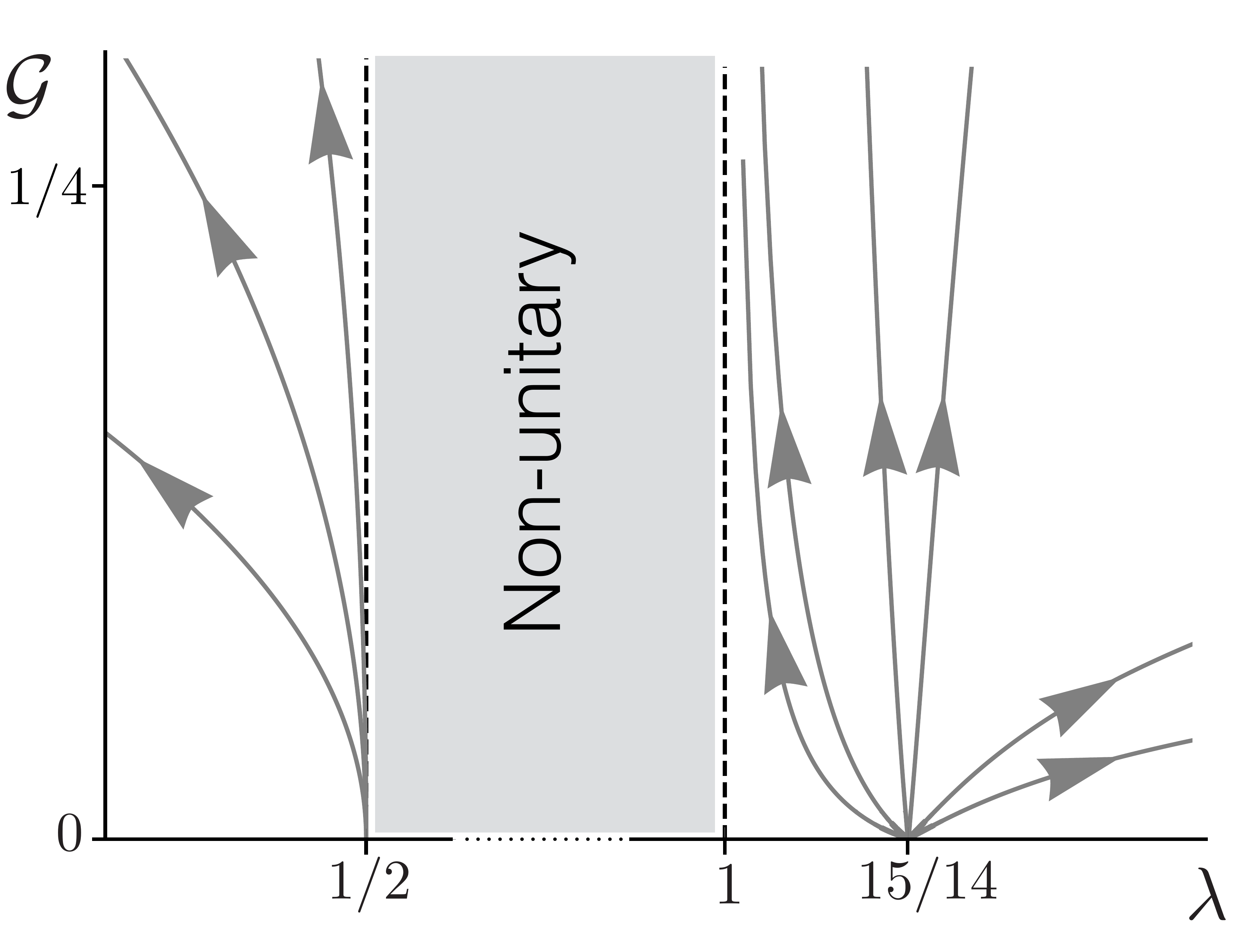}}
\caption{RG flow of the couplings in
  $(2+1)$-dimensional Ho\v rava gravity. The arrows show the direction of
  the flow towards the infrared.
\label{flow}}
\vspace{-0.2cm}
\end{figure}

The structure of the RG flow in the regions required by unitarity
\cite{Barvinsky:2015kil}, $\{\l<1/2\}\cup\{\l>1\}$, is shown in
Fig.~\ref{flow}. 
The theory possesses two UV fixed points: 
\begin{align}
\label{FP}
&(\l,{\cal G})=\left(\frac{1}{2},0\right)\quad \mathrm{and}\quad 
(\l,{\cal G})=\left(\frac{15}{14},0\right)\;.
\end{align}
The first fixed point is located at the boundary of the allowed
region. The limit $\l\to 1/2$, taken at fixed value of ${\cal G}$, is
strongly coupled, as it is clear from the divergence in
\eqref{eq:betaofg}. However, the RG flow $\lambda\to 1/2$ is
accompanied by the vanishing of ${\cal G}$. The structure of
$\beta_{\cal G}$ in 
\eqref{eq:betaofg} suggests that the actual expansion parameter in this
limit is\footnote{The analysis of the interaction 
terms in the action
  confirms that this combination controls the coupling strength when $\lambda\to1/2$.} 
$\tilde{\cal
G}={\cal G}/\sqrt{1-2\l}$, with the
$\beta$-function
\begin{align}
\label{gtildeRG}
\beta_{\tilde {\cal G}}=-\frac{(1-2\l)^2}{64\pi (1-\l)^{3/2}}\;\tilde {\cal G}^2\;.
\end{align}
This $\b$-function vanishes at $\l\to 1/2$, so that $\tilde {\cal G}$
freezes at a constant value in the UV. In other words,  at the
one-loop level there is  
a family of UV fixed points parameterized by the 
asymptotic value of $\tilde {\cal G}$. The status of this fixed-point family can be clarified only by taking into account contributions from higher order and matter loops.

Remarkably, the second UV fixed point in (\ref{FP}) is regular
and asymptotically free. 
In the infrared (IR), the RG trajectories either
go to $\l\to +\infty,~{\cal G}\to +\infty$, or to $\l\to
1^+,~{\cal G}\to+\infty$. The latter behavior is intriguing as it naively
corresponds to the relativistic limit of the theory. However, to
decide whether the theory really flows or not to GR requires a
non-perturbative analysis as in the IR the system enters into the
strong-coupling regime, which is typical for asymptotically free
theories\footnote{If this behavior persists in higher dimensions, 
the strong coupling in the IR will be naturally cut off by relevant
deformations, which do not exist in $2+1$ dimensions,
cf. \eqref{eq:action_2}.}. 

It is worth comparing our results to those of
Ref.~\cite{Benedetti:2013pya} where the  
gravitational degrees 
of freedom were truncated to the conformal mode. While the family of
fixed points at $\lambda\to 1/2$ is present also in the truncated
model, the fixed point at $\lambda =15/14$ exists
only in the full theory. A more important difference is that the
asymptotic freedom of the gravitational coupling reported in
\cite{Benedetti:2013pya} occurrs at {\em negative} values of $G$ 
--- the choice required by unitarity of the truncated
model at $\lambda>1/2$. On the other hand, the full theory is both
unitary and asymptotically free at postive $G$ and $\l>1$.

\paragraph*{{\bf Conclusions} -}
The main result presented in this letter are the $\beta$-functions
\eqref{eq:betas} for the essential
coupling constants of projectable Ho\v rava gravity in $2+1$
dimensions. Their associated flow is shown in Fig.~\ref{flow}.  
The RG flow possesses an asymptotically-free fixed point in the UV,
which establishes this model as a $2+1$ dimensional perturbatively
UV-complete theory 
with non-trivial propagating gravitational degrees of freedom. 
This not only makes it a theory of quantum gravity in $2+1$ dimensions
but also a suitable toy model to address fundamental aspects of quantum
gravity in the more realistic $(3+1)$-dimensional case.

\paragraph*{{\bf Acknowledgements} -}
We are grateful to Renate Loll and Frank Saueressig for discussions.
M. H-V. is grateful to Jos Vermaseren for his help with FORM during
early stages of this work. C.S. is grateful for the hospitality of the
CERN theory division. This work was 
supported by the RFBR grant No.17-02-00651 (A.B. and S.S.), the Tomsk State
University Competitiveness Improvement Program (A.B.),  
the Tomalla Foundation (M.H.-V.) and the Swiss
National Science Foundation (S.S.).


\bibliography{biblio_letter2}{}

\begin{thebibliography}{26}%
\makeatletter
\providecommand \@ifxundefined [1]{%
 \@ifx{#1\undefined}
}%
\providecommand \@ifnum [1]{%
 \ifnum #1\expandafter \@firstoftwo
 \else \expandafter \@secondoftwo
 \fi
}%
\providecommand \@ifx [1]{%
 \ifx #1\expandafter \@firstoftwo
 \else \expandafter \@secondoftwo
 \fi
}%
\providecommand \natexlab [1]{#1}%
\providecommand \enquote  [1]{``#1''}%
\providecommand \bibnamefont  [1]{#1}%
\providecommand \bibfnamefont [1]{#1}%
\providecommand \citenamefont [1]{#1}%
\providecommand \href@noop [0]{\@secondoftwo}%
\providecommand \href [0]{\begingroup \@sanitize@url \@href}%
\providecommand \@href[1]{\@@startlink{#1}\@@href}%
\providecommand \@@href[1]{\endgroup#1\@@endlink}%
\providecommand \@sanitize@url [0]{\catcode `\\12\catcode `\$12\catcode
  `\&12\catcode `\#12\catcode `\^12\catcode `\_12\catcode `\%12\relax}%
\providecommand \@@startlink[1]{}%
\providecommand \@@endlink[0]{}%
\providecommand \url  [0]{\begingroup\@sanitize@url \@url }%
\providecommand \@url [1]{\endgroup\@href {#1}{\urlprefix }}%
\providecommand \urlprefix  [0]{URL }%
\providecommand \Eprint [0]{\href }%
\providecommand \doibase [0]{http://dx.doi.org/}%
\providecommand \selectlanguage [0]{\@gobble}%
\providecommand \bibinfo  [0]{\@secondoftwo}%
\providecommand \bibfield  [0]{\@secondoftwo}%
\providecommand \translation [1]{[#1]}%
\providecommand \BibitemOpen [0]{}%
\providecommand \bibitemStop [0]{}%
\providecommand \bibitemNoStop [0]{.\EOS\space}%
\providecommand \EOS [0]{\spacefactor3000\relax}%
\providecommand \BibitemShut  [1]{\csname bibitem#1\endcsname}%
\let\auto@bib@innerbib\@empty
\bibitem [{\citenamefont {Stelle}(1977)}]{Stelle:1976gc}%
  \BibitemOpen
  \bibfield  {author} {\bibinfo {author} {\bibfnamefont {K.~S.}\ \bibnamefont
  {Stelle}},\ }\href {\doibase 10.1103/PhysRevD.16.953} {\bibfield  {journal}
  {\bibinfo  {journal} {Phys. Rev.}\ }\textbf {\bibinfo {volume} {D16}},\
  \bibinfo {pages} {953} (\bibinfo {year} {1977})}\BibitemShut {NoStop}%
\bibitem [{\citenamefont {Fradkin}\ and\ \citenamefont
  {Tseytlin}(1981)}]{Fradkin:1981hx}%
  \BibitemOpen
  \bibfield  {author} {\bibinfo {author} {\bibfnamefont {E.~S.}\ \bibnamefont
  {Fradkin}}\ and\ \bibinfo {author} {\bibfnamefont {A.~A.}\ \bibnamefont
  {Tseytlin}},\ }\href {\doibase 10.1016/0370-2693(81)90702-4} {\bibfield
  {journal} {\bibinfo  {journal} {Phys. Lett.}\ }\textbf {\bibinfo {volume}
  {104B}},\ \bibinfo {pages} {377} (\bibinfo {year} {1981})}\BibitemShut
  {NoStop}%
\bibitem [{\citenamefont {Avramidi}\ and\ \citenamefont
  {Barvinsky}(1985)}]{Avramidi:1985ki}%
  \BibitemOpen
  \bibfield  {author} {\bibinfo {author} {\bibfnamefont {I.~G.}\ \bibnamefont
  {Avramidi}}\ and\ \bibinfo {author} {\bibfnamefont {A.~O.}\ \bibnamefont
  {Barvinsky}},\ }\href {\doibase 10.1016/0370-2693(85)90248-5} {\bibfield
  {journal} {\bibinfo  {journal} {Phys. Lett.}\ }\textbf {\bibinfo {volume}
  {B159}},\ \bibinfo {pages} {269} (\bibinfo {year} {1985})}\BibitemShut
  {NoStop}%
\bibitem [{\citenamefont {Horava}(2009)}]{Horava:2009uw}%
  \BibitemOpen
  \bibfield  {author} {\bibinfo {author} {\bibfnamefont {P.}~\bibnamefont
  {Horava}},\ }\href {\doibase 10.1103/PhysRevD.79.084008} {\bibfield
  {journal} {\bibinfo  {journal} {Phys. Rev.}\ }\textbf {\bibinfo {volume}
  {D79}},\ \bibinfo {pages} {084008} (\bibinfo {year} {2009})},\ \Eprint
  {http://arxiv.org/abs/0901.3775} {arXiv:0901.3775 [hep-th]} \BibitemShut
  {NoStop}%
\bibitem [{\citenamefont {Liberati}(2013)}]{Liberati:2013xla}%
  \BibitemOpen
  \bibfield  {author} {\bibinfo {author} {\bibfnamefont {S.}~\bibnamefont
  {Liberati}},\ }\href {\doibase 10.1088/0264-9381/30/13/133001} {\bibfield
  {journal} {\bibinfo  {journal} {Class. Quant. Grav.}\ }\textbf {\bibinfo
  {volume} {30}},\ \bibinfo {pages} {133001} (\bibinfo {year} {2013})},\
  \Eprint {http://arxiv.org/abs/1304.5795} {arXiv:1304.5795 [gr-qc]}
  \BibitemShut {NoStop}%
\bibitem [{\citenamefont {Blas}\ \emph {et~al.}(2010)\citenamefont {Blas},
  \citenamefont {Pujolas},\ and\ \citenamefont {Sibiryakov}}]{Blas:2009qj}%
  \BibitemOpen
  \bibfield  {author} {\bibinfo {author} {\bibfnamefont {D.}~\bibnamefont
  {Blas}}, \bibinfo {author} {\bibfnamefont {O.}~\bibnamefont {Pujolas}}, \
  and\ \bibinfo {author} {\bibfnamefont {S.}~\bibnamefont {Sibiryakov}},\
  }\href {\doibase 10.1103/PhysRevLett.104.181302} {\bibfield  {journal}
  {\bibinfo  {journal} {Phys. Rev. Lett.}\ }\textbf {\bibinfo {volume} {104}},\
  \bibinfo {pages} {181302} (\bibinfo {year} {2010})},\ \Eprint
  {http://arxiv.org/abs/0909.3525} {arXiv:0909.3525 [hep-th]} \BibitemShut
  {NoStop}%
\bibitem [{\citenamefont {Blas}\ and\ \citenamefont
  {Lim}(2015)}]{Blas:2014aca}%
  \BibitemOpen
  \bibfield  {author} {\bibinfo {author} {\bibfnamefont {D.}~\bibnamefont
  {Blas}}\ and\ \bibinfo {author} {\bibfnamefont {E.}~\bibnamefont {Lim}},\
  }\href {\doibase 10.1142/S0218271814430093} {\bibfield  {journal} {\bibinfo
  {journal} {Int. J. Mod. Phys.}\ }\textbf {\bibinfo {volume} {D23}},\ \bibinfo
  {pages} {1443009} (\bibinfo {year} {2015})},\ \Eprint
  {http://arxiv.org/abs/1412.4828} {arXiv:1412.4828 [gr-qc]} \BibitemShut
  {NoStop}%
\bibitem [{\citenamefont {Janiszewski}\ and\ \citenamefont
  {Karch}(2013)}]{Janiszewski:2012nb}%
  \BibitemOpen
  \bibfield  {author} {\bibinfo {author} {\bibfnamefont {S.}~\bibnamefont
  {Janiszewski}}\ and\ \bibinfo {author} {\bibfnamefont {A.}~\bibnamefont
  {Karch}},\ }\href {\doibase 10.1007/JHEP02(2013)123} {\bibfield  {journal}
  {\bibinfo  {journal} {JHEP}\ }\textbf {\bibinfo {volume} {02}},\ \bibinfo
  {pages} {123} (\bibinfo {year} {2013})},\ \Eprint
  {http://arxiv.org/abs/1211.0005} {arXiv:1211.0005 [hep-th]} \BibitemShut
  {NoStop}%
\bibitem [{\citenamefont {Griffin}\ \emph {et~al.}(2013)\citenamefont
  {Griffin}, \citenamefont {Horava},\ and\ \citenamefont
  {Melby-Thompson}}]{Griffin:2012qx}%
  \BibitemOpen
  \bibfield  {author} {\bibinfo {author} {\bibfnamefont {T.}~\bibnamefont
  {Griffin}}, \bibinfo {author} {\bibfnamefont {P.}~\bibnamefont {Horava}}, \
  and\ \bibinfo {author} {\bibfnamefont {C.~M.}\ \bibnamefont
  {Melby-Thompson}},\ }\href {\doibase 10.1103/PhysRevLett.110.081602}
  {\bibfield  {journal} {\bibinfo  {journal} {Phys. Rev. Lett.}\ }\textbf
  {\bibinfo {volume} {110}},\ \bibinfo {pages} {081602} (\bibinfo {year}
  {2013})},\ \Eprint {http://arxiv.org/abs/1211.4872} {arXiv:1211.4872
  [hep-th]} \BibitemShut {NoStop}%
\bibitem [{\citenamefont {Barvinsky}\ \emph {et~al.}(2016)\citenamefont
  {Barvinsky}, \citenamefont {Blas}, \citenamefont {Herrero-Valea},
  \citenamefont {Sibiryakov},\ and\ \citenamefont
  {Steinwachs}}]{Barvinsky:2015kil}%
  \BibitemOpen
  \bibfield  {author} {\bibinfo {author} {\bibfnamefont {A.~O.}\ \bibnamefont
  {Barvinsky}}, \bibinfo {author} {\bibfnamefont {D.}~\bibnamefont {Blas}},
  \bibinfo {author} {\bibfnamefont {M.}~\bibnamefont {Herrero-Valea}}, \bibinfo
  {author} {\bibfnamefont {S.~M.}\ \bibnamefont {Sibiryakov}}, \ and\ \bibinfo
  {author} {\bibfnamefont {C.~F.}\ \bibnamefont {Steinwachs}},\ }\href
  {\doibase 10.1103/PhysRevD.93.064022} {\bibfield  {journal} {\bibinfo
  {journal} {Phys. Rev.}\ }\textbf {\bibinfo {volume} {D93}},\ \bibinfo {pages}
  {064022} (\bibinfo {year} {2016})},\ \Eprint
  {http://arxiv.org/abs/1512.02250} {arXiv:1512.02250 [hep-th]} \BibitemShut
  {NoStop}%
\bibitem [{\citenamefont {Barvinsky}\ \emph {et~al.}(2017)\citenamefont
  {Barvinsky}, \citenamefont {Blas}, \citenamefont {Herrero-Valea},
  \citenamefont {Sibiryakov},\ and\ \citenamefont
  {Steinwachs}}]{Barvinsky:2017zlx}%
  \BibitemOpen
  \bibfield  {author} {\bibinfo {author} {\bibfnamefont {A.~O.}\ \bibnamefont
  {Barvinsky}}, \bibinfo {author} {\bibfnamefont {D.}~\bibnamefont {Blas}},
  \bibinfo {author} {\bibfnamefont {M.}~\bibnamefont {Herrero-Valea}}, \bibinfo
  {author} {\bibfnamefont {S.~M.}\ \bibnamefont {Sibiryakov}}, \ and\ \bibinfo
  {author} {\bibfnamefont {C.~F.}\ \bibnamefont {Steinwachs}},\ }\href@noop {}
  {\enquote {\bibinfo {title} {{Renormalization of gauge theories in the
  background-field approach}},}\ } (\bibinfo {year} {2017}),\ \Eprint
  {http://arxiv.org/abs/1705.03480} {arXiv:1705.03480 [hep-th]} \BibitemShut
  {NoStop}%
\bibitem [{\citenamefont {Benedetti}\ and\ \citenamefont
  {Guarnieri}(2014)}]{Benedetti:2013pya}%
  \BibitemOpen
  \bibfield  {author} {\bibinfo {author} {\bibfnamefont {D.}~\bibnamefont
  {Benedetti}}\ and\ \bibinfo {author} {\bibfnamefont {F.}~\bibnamefont
  {Guarnieri}},\ }\href {\doibase 10.1007/JHEP03(2014)078} {\bibfield
  {journal} {\bibinfo  {journal} {JHEP}\ }\textbf {\bibinfo {volume} {03}},\
  \bibinfo {pages} {078} (\bibinfo {year} {2014})},\ \Eprint
  {http://arxiv.org/abs/1311.6253} {arXiv:1311.6253 [hep-th]} \BibitemShut
  {NoStop}%
\bibitem [{\citenamefont {D'Odorico}\ \emph {et~al.}(2014)\citenamefont
  {D'Odorico}, \citenamefont {Saueressig},\ and\ \citenamefont
  {Schutten}}]{DOdorico:2014tyh}%
  \BibitemOpen
  \bibfield  {author} {\bibinfo {author} {\bibfnamefont {G.}~\bibnamefont
  {D'Odorico}}, \bibinfo {author} {\bibfnamefont {F.}~\bibnamefont
  {Saueressig}}, \ and\ \bibinfo {author} {\bibfnamefont {M.}~\bibnamefont
  {Schutten}},\ }\href {\doibase 10.1103/PhysRevLett.113.171101} {\bibfield
  {journal} {\bibinfo  {journal} {Phys. Rev. Lett.}\ }\textbf {\bibinfo
  {volume} {113}},\ \bibinfo {pages} {171101} (\bibinfo {year} {2014})},\
  \Eprint {http://arxiv.org/abs/1406.4366} {arXiv:1406.4366 [gr-qc]}
  \BibitemShut {NoStop}%
\bibitem [{\citenamefont {Sotiriou}\ \emph {et~al.}(2011)\citenamefont
  {Sotiriou}, \citenamefont {Visser},\ and\ \citenamefont
  {Weinfurtner}}]{Sotiriou:2011dr}%
  \BibitemOpen
  \bibfield  {author} {\bibinfo {author} {\bibfnamefont {T.~P.}\ \bibnamefont
  {Sotiriou}}, \bibinfo {author} {\bibfnamefont {M.}~\bibnamefont {Visser}}, \
  and\ \bibinfo {author} {\bibfnamefont {S.}~\bibnamefont {Weinfurtner}},\
  }\href {\doibase 10.1103/PhysRevD.83.124021} {\bibfield  {journal} {\bibinfo
  {journal} {Phys. Rev.}\ }\textbf {\bibinfo {volume} {D83}},\ \bibinfo {pages}
  {124021} (\bibinfo {year} {2011})},\ \Eprint {http://arxiv.org/abs/1103.3013}
  {arXiv:1103.3013 [hep-th]} \BibitemShut {NoStop}%
\bibitem [{\citenamefont {Sotiriou}\ \emph
  {et~al.}(2009{\natexlab{a}})\citenamefont {Sotiriou}, \citenamefont
  {Visser},\ and\ \citenamefont {Weinfurtner}}]{Sotiriou:2009gy}%
  \BibitemOpen
  \bibfield  {author} {\bibinfo {author} {\bibfnamefont {T.~P.}\ \bibnamefont
  {Sotiriou}}, \bibinfo {author} {\bibfnamefont {M.}~\bibnamefont {Visser}}, \
  and\ \bibinfo {author} {\bibfnamefont {S.}~\bibnamefont {Weinfurtner}},\
  }\href {\doibase 10.1103/PhysRevLett.102.251601} {\bibfield  {journal}
  {\bibinfo  {journal} {Phys. Rev. Lett.}\ }\textbf {\bibinfo {volume} {102}},\
  \bibinfo {pages} {251601} (\bibinfo {year} {2009}{\natexlab{a}})},\ \Eprint
  {http://arxiv.org/abs/0904.4464} {arXiv:0904.4464 [hep-th]} \BibitemShut
  {NoStop}%
\bibitem [{\citenamefont {Sotiriou}\ \emph
  {et~al.}(2009{\natexlab{b}})\citenamefont {Sotiriou}, \citenamefont
  {Visser},\ and\ \citenamefont {Weinfurtner}}]{Sotiriou:2009bx}%
  \BibitemOpen
  \bibfield  {author} {\bibinfo {author} {\bibfnamefont {T.~P.}\ \bibnamefont
  {Sotiriou}}, \bibinfo {author} {\bibfnamefont {M.}~\bibnamefont {Visser}}, \
  and\ \bibinfo {author} {\bibfnamefont {S.}~\bibnamefont {Weinfurtner}},\
  }\href {\doibase 10.1088/1126-6708/2009/10/033} {\bibfield  {journal}
  {\bibinfo  {journal} {JHEP}\ }\textbf {\bibinfo {volume} {10}},\ \bibinfo
  {pages} {033} (\bibinfo {year} {2009}{\natexlab{b}})},\ \Eprint
  {http://arxiv.org/abs/0905.2798} {arXiv:0905.2798 [hep-th]} \BibitemShut
  {NoStop}%
\bibitem [{\citenamefont {Blas}\ \emph {et~al.}(2011)\citenamefont {Blas},
  \citenamefont {Pujolas},\ and\ \citenamefont {Sibiryakov}}]{Blas:2010hb}%
  \BibitemOpen
  \bibfield  {author} {\bibinfo {author} {\bibfnamefont {D.}~\bibnamefont
  {Blas}}, \bibinfo {author} {\bibfnamefont {O.}~\bibnamefont {Pujolas}}, \
  and\ \bibinfo {author} {\bibfnamefont {S.}~\bibnamefont {Sibiryakov}},\
  }\href {\doibase 10.1007/JHEP04(2011)018} {\bibfield  {journal} {\bibinfo
  {journal} {JHEP}\ }\textbf {\bibinfo {volume} {04}},\ \bibinfo {pages} {018}
  (\bibinfo {year} {2011})},\ \Eprint {http://arxiv.org/abs/1007.3503}
  {arXiv:1007.3503 [hep-th]} \BibitemShut {NoStop}%
\bibitem [{\citenamefont {Abbott}(1982)}]{Abbott:1981ke}%
  \BibitemOpen
  \bibfield  {author} {\bibinfo {author} {\bibfnamefont {L.~F.}\ \bibnamefont
  {Abbott}},\ }\href@noop {} {\bibfield  {journal} {\bibinfo  {journal} {Acta
  Phys. Polon.}\ }\textbf {\bibinfo {volume} {B13}},\ \bibinfo {pages} {33}
  (\bibinfo {year} {1982})}\BibitemShut {NoStop}%
\bibitem [{\citenamefont {Griffin}\ \emph {et~al.}(2017)\citenamefont
  {Griffin}, \citenamefont {Grosvenor}, \citenamefont {Melby-Thompson},\ and\
  \citenamefont {Yan}}]{Griffin:2017wvh}%
  \BibitemOpen
  \bibfield  {author} {\bibinfo {author} {\bibfnamefont {T.}~\bibnamefont
  {Griffin}}, \bibinfo {author} {\bibfnamefont {K.~T.}\ \bibnamefont
  {Grosvenor}}, \bibinfo {author} {\bibfnamefont {C.~M.}\ \bibnamefont
  {Melby-Thompson}}, \ and\ \bibinfo {author} {\bibfnamefont {Z.}~\bibnamefont
  {Yan}},\ }\href {\doibase 10.1007/JHEP06(2017)004} {\bibfield  {journal}
  {\bibinfo  {journal} {JHEP}\ }\textbf {\bibinfo {volume} {06}},\ \bibinfo
  {pages} {004} (\bibinfo {year} {2017})},\ \Eprint
  {http://arxiv.org/abs/1701.08173} {arXiv:1701.08173 [hep-th]} \BibitemShut
  {NoStop}%
\bibitem [{\citenamefont {DeWitt}(1967)}]{DeWitt:1967ub}%
  \BibitemOpen
  \bibfield  {author} {\bibinfo {author} {\bibfnamefont {B.~S.}\ \bibnamefont
  {DeWitt}},\ }\href {\doibase 10.1103/PhysRev.162.1195} {\bibfield  {journal}
  {\bibinfo  {journal} {Phys. Rev.}\ }\textbf {\bibinfo {volume} {162}},\
  \bibinfo {pages} {1195} (\bibinfo {year} {1967})}\BibitemShut {NoStop}%
\bibitem [{\citenamefont {Kallosh}(1974)}]{Kallosh:1974yh}%
  \BibitemOpen
  \bibfield  {author} {\bibinfo {author} {\bibfnamefont {R.~E.}\ \bibnamefont
  {Kallosh}},\ }\href {\doibase 10.1016/0550-3213(74)90284-3} {\bibfield
  {journal} {\bibinfo  {journal} {Nucl. Phys.}\ }\textbf {\bibinfo {volume}
  {B78}},\ \bibinfo {pages} {293} (\bibinfo {year} {1974})}\BibitemShut
  {NoStop}%
\bibitem [{\citenamefont {Anselmi}(2009)}]{Anselmi:2008bq}%
  \BibitemOpen
  \bibfield  {author} {\bibinfo {author} {\bibfnamefont {D.}~\bibnamefont
  {Anselmi}},\ }\href {\doibase 10.1016/j.aop.2008.12.005} {\bibfield
  {journal} {\bibinfo  {journal} {Annals Phys.}\ }\textbf {\bibinfo {volume}
  {324}},\ \bibinfo {pages} {874} (\bibinfo {year} {2009})},\ \Eprint
  {http://arxiv.org/abs/0808.3470} {arXiv:0808.3470 [hep-th]} \BibitemShut
  {NoStop}%
\bibitem [{\citenamefont {Brizuela}\ \emph {et~al.}(2009)\citenamefont
  {Brizuela}, \citenamefont {Martin-Garcia},\ and\ \citenamefont
  {Mena~Marugan}}]{Brizuela:2008ra}%
  \BibitemOpen
  \bibfield  {author} {\bibinfo {author} {\bibfnamefont {D.}~\bibnamefont
  {Brizuela}}, \bibinfo {author} {\bibfnamefont {J.~M.}\ \bibnamefont
  {Martin-Garcia}}, \ and\ \bibinfo {author} {\bibfnamefont {G.~A.}\
  \bibnamefont {Mena~Marugan}},\ }\href {\doibase 10.1007/s10714-009-0773-2}
  {\bibfield  {journal} {\bibinfo  {journal} {Gen. Rel. Grav.}\ }\textbf
  {\bibinfo {volume} {41}},\ \bibinfo {pages} {2415} (\bibinfo {year}
  {2009})},\ \Eprint {http://arxiv.org/abs/0807.0824} {arXiv:0807.0824 [gr-qc]}
  \BibitemShut {NoStop}%
\bibitem [{\citenamefont {Ueda}\ and\ \citenamefont
  {Vermaseren}(2014)}]{Ueda:2014sya}%
  \BibitemOpen
  \bibfield  {author} {\bibinfo {author} {\bibfnamefont {T.}~\bibnamefont
  {Ueda}}\ and\ \bibinfo {author} {\bibfnamefont {J.}~\bibnamefont
  {Vermaseren}},\ }\bibfield  {booktitle} {\emph {\bibinfo {booktitle}
  {{Proceedings, 15th International Workshop on Advanced Computing and Analysis
  Techniques in Physics Research (ACAT 2013): Beijing, China, May 16-21,
  2013}}},\ }\href {\doibase 10.1088/1742-6596/523/1/012047} {\bibfield
  {journal} {\bibinfo  {journal} {J. Phys. Conf. Ser.}\ }\textbf {\bibinfo
  {volume} {523}},\ \bibinfo {pages} {012047} (\bibinfo {year}
  {2014})}\BibitemShut {NoStop}%
\bibitem [{\citenamefont {Liao}(1996)}]{Liao:1994fp}%
  \BibitemOpen
  \bibfield  {author} {\bibinfo {author} {\bibfnamefont {S.-B.}\ \bibnamefont
  {Liao}},\ }\href {\doibase 10.1103/PhysRevD.53.2020} {\bibfield  {journal}
  {\bibinfo  {journal} {Phys. Rev.}\ }\textbf {\bibinfo {volume} {D53}},\
  \bibinfo {pages} {2020} (\bibinfo {year} {1996})},\ \Eprint
  {http://arxiv.org/abs/hep-th/9501124} {arXiv:hep-th/9501124 [hep-th]}
  \BibitemShut {NoStop}%
\bibitem [{\citenamefont {Barvinsky}\ \emph {et~al.}()\citenamefont
  {Barvinsky}, \citenamefont {Blas}, \citenamefont {Herrero-Valea},
  \citenamefont {Sibiryakov},\ and\ \citenamefont {Steinwachs}}]{future}%
  \BibitemOpen
  \bibfield  {author} {\bibinfo {author} {\bibfnamefont {A.~O.}\ \bibnamefont
  {Barvinsky}}, \bibinfo {author} {\bibfnamefont {D.}~\bibnamefont {Blas}},
  \bibinfo {author} {\bibfnamefont {M.}~\bibnamefont {Herrero-Valea}}, \bibinfo
  {author} {\bibfnamefont {S.~M.}\ \bibnamefont {Sibiryakov}}, \ and\ \bibinfo
  {author} {\bibfnamefont {C.~F.}\ \bibnamefont {Steinwachs}},\ }\href@noop {}
  {\bibinfo  {journal} {{\it in preparation}}\ }\BibitemShut {NoStop}%
\end{thebibliography}%

\end{document}